\documentclass{bio}
\usepackage{dcolumn, color}
\usepackage{amsmath, modamssymb, modamsfonts, modamsthm, modbm}
\usepackage{times, modlastpage}
\usepackage{graphicx,lscape}
\usepackage{modnatbib}
\numberwithin{equation}{section}
\pyear{2009}
\jno{}

\begin{document}
\frenchspacing

\title{Contact intervals, survival analysis of epidemic data, and estimation of $R_0$}

\author{EBEN KENAH$^\ast$}
\address{Departments of Biostatistics and Global Health}
\address{University of Washington, Seattle, WA 98105, USA}
\email{eek4@u.washington.edu}
\maketitle

\pagestyle{headings}
\markboth{E. Kenah}{Contact intervals, survival analysis of epidemic data, and estimation of $R_0$}

\begin{abstract}
{We argue that the time from the onset of infectiousness to infectious contact, which we call the \textit{contact interval}, is a better basis for inference in epidemic data than the generation or serial interval.  Since contact intervals can be right-censored, survival analysis is the natural approach to estimation.  Estimates of the contact interval distribution can be used to estimate $R_0$ in both mass-action and network-based models.}
\vspace*{-6pt}
{Basic reproductive number ($R_0$); Epidemic data; Generation intervals; Survival analysis}
\end{abstract}

\section{Introduction}
Infectious disease remains one of the greatest threats to human health and commerce, and the analysis of epidemic data is one of the most important applications of statistics in public health.  Some of the most important questions involve the basic reproductive number, $R_0$, the number of secondary infections caused by a typical infectious person in the early stages of an epidemic \citep{DiekmannHeesterbeek}.  Higher values of $R_0$ indicate that an epidemic will be larger and harder to control.  The effects of interventions and the depletion of the susceptible population can be captured with the effective reproductive number $R(t)$, which is the number of secondary infections caused by a typical person infected at time $t$.

The \textit{generation interval} of an infectious disease is the time between the infection of a secondary case and the infection of his or her infector.  The \textit{serial interval} is the time between the symptom onset of a secondary case and the symptom onset of his or her infector.  The generation and serial interval distributions are often considered characteristic features of an infectious disease \citep{Fine}.  For a given $R_0$, a shorter mean serial or generation interval implies faster spread of the epidemic.  

Usually, generation intervals are times between unobserved events.  Serial intervals, which are times between observed events, are often used instead.  Recent analyses of several past, emerging, and potentially emerging infectious diseases have been based on serial interval distributions, including the 1918 influenza \citep{Mills}, Severe Acute Respiratory Syndrome (SARS) \citep{LipsitchSARS, WallingaTeunis}, pandemic influenza A(H1N1) \citep{FraserH1N1, McBryde}, and avian influenza \citep{Ferguson1, Ferguson2}.  Three methods form the basis of these applications.  With a measurement of the exponential growth rate at the beginning of an epidemic and a known serial interval distribution, $R_0$ can be estimated via the Lotka-Euler equation \citep{DiekmannHeesterbeek, Svensson, WallingaLipsitch, RobertsHeesterbeek}.  Two other methods use the time series of symptom onset times, assuming that all infections are symptomatic and observed.  \citet{WallingaTeunis} estimate $R(t)$ given a known serial interval distribution.  Their approach has been adapted by other researchers \citep{Cauchemez}, often supplemented with serial-interval observations from contact-tracing data.  \citet{WhitePagano} jointly estimate $R_0$ and the serial interval distribution using a branching-process approximation to the initial spread of infection, assuming the number of secondary cases generated by each infectious person has a Poisson distribution with mean $R_0$.

There are several problems with estimators based on generation or serial intervals in the context of an emerging infection.  The Lotka-Euler and Wallinga-Teunis estimators rely on a previously known generation serial interval distribution.  The Wallinga-Teunis and White-Pagano estimators assume that all serial intervals are independent and identically distributed, which occurs only if the incubation and infectious periods are constant.  All three of these estimators assume a stable serial interval distribution, which limits their use to the early spread of infection.  When multiple infectious persons compete to infect a given susceptible, the infector is the one who first makes infectious contact.  Thus, the mean generation and serial intervals contract as the prevalence of infection increases either locally (e.g., within households) or globally \citep{Svensson, Kenah3}.

In this paper, we outline an alternative analysis of epidemic data that applies methods from survival analysis to \textit{contact intervals}.  Informally, the contact interval from an infectious person $i$ to a susceptible person $j$ is the time between the onset of infectiousness in $i$ and the first infectious contact from $i$ to $j$, where we define infectious contact to be a contact sufficient to infect a susceptible individual.  This interval will be right-censored if $j$ is infected by someone else prior to infectious contact from $i$ or if $i$ recovers from infection before making infectious contact with $j$.  The contact interval is similar to the generation interval, except that its definition is not limited to contacts that actually cause infection and it begins with the onset of infectiousness rather than infection.  

Here, we focus on the analysis of completely-observed ``Susceptible-Exposed-Infectious-Recovered'' (SEIR) epidemics.  The SEIR framework applies to acute, immunizing diseases that spread from person to person, such as measles, influenza, smallpox, and polio.  We also assume that the epidemic is completely observed, so all cases are detected and their times of infection, onset of infectiousness, and recovery are observed.  Most epidemics are only partially observed, so we plan to explore the analysis of more realistic data sets in future papers.  However, it is best viewed as a missing data problem, which requires that the methods for complete data be established.

In Section 2, we define a general stochastic SEIR epidemic model and show that survival likelihoods for a vector $\theta$ of contact interval distribution parameters have score processes that are zero-mean martingales at the true parameter $\theta_0$.  In Section 3, we show how estimates of the contact interval distribution can be used to estimate $R_0$ in mass-action and network-based models.  In Section 4, we evaluate the performance of these methods in simulated epidemic data and show that assumptions about the underlying contact process play a crucial role in accurate statistical inference.  In Section 5, we discuss the advantages and limitations of survival methods in epidemic data analysis.

\section{Methods}
In this section, we show that the score processes from survival likelihoods for epidemic data can be written as stochastic integrals with respect to zero-mean martingales.  We develop our methods in three stages.  First, we describe the underlying stochastic SEIR model and the observed data.  Second, we imagine that we observe who-infected-whom and derive counting-process martingales for an ordered pair $ij$ and for a fixed susceptible $j$.  Finally, we consider the situation where we do not observe who-infected-whom and derive counting-process martingales for a fixed susceptible $j$ and for the complete observed data.  Our sources for the underlying theory are \citet{KalbfleischPrentice} and \citet{Serfling}.

\subsection{Stochastic SEIR model and observed data}
Consider a stochastic ``Susceptible-Exposed-Infectious-Removed'' (SEIR) model in a closed population of $n$ individuals assigned indices $1,\ldots, n$.  Each person $i$ moves from S to E at his or her \textit{infection time} $t_i$, with $t_i = \infty$ if $i$ is never infected.  After infection, $i$ begins a \textit{latent period} of length $\varepsilon_i$ during which he or she is infected but not infectious.  At time $t_i + \varepsilon_i$, $i$ moves from E to I, beginning an \textit{infectious period} of length $\iota_i$.  At time $t_i + r_i$, $i$ recovers from infection and moves from I to R, where the \textit{recovery period} $r_i = \varepsilon_i + \iota_i$ is the total time between infection and removal.  Once in R, $i$ can no longer infect other persons or be infected.  The latent period is a nonnegative random variable, the infectious and recovery periods are strictly positive random variables, and the recovery period is finite with probability one.

After becoming infectious at time $t_i + \varepsilon_i$, person $i$ makes infectious contact with person $j\neq i$ at their \textit{infectious contact time} $t_{ij} = t_i + \varepsilon_i + \tau_{ij}^*$, where the \textit{infectious contact interval} $\tau_{ij}^*$ is a strictly positive random variable with $\tau_{ij}^* = \infty$ if infectious contact never occurs.  Since infectious contact must occur while $i$ is infectious or never, $\tau_{ij}^*\in(0, \iota_i]$ or $\tau_{ij}^* = \infty$.  We define infectious contact to be sufficient to cause infection in a susceptible person, so $t_j\leq t_{ij}$ with equality if and only if $j$ is susceptible at time $t_{ij}$.

An epidemic begins with one or more persons infected from outside the population, which we call \textit{imported infections}.  For simplicity, we assume that epidemics begin with one or more imported infections at time $t = 0$ and there are no other imported infections.

\paragraph{Contact intervals} For each ordered pair $ij$, let $C_{ij} = 1$ if infectious contact from $i$ to $j$ is possible and $C_{ij} = 0$ otherwise.  We assume that the infectious contact interval $\tau_{ij}^*$ is generated in the following way: A \textit{contact interval} $\tau_{ij}$ is drawn from a distribution with hazard function $\lambda_{ij}(\tau)$.  If $\tau_{ij}\leq \iota_i$ and $C_{ij} = 1$, then $\tau_{ij}^* = \tau_{ij}$.  Otherwise $\tau_{ij}^* = \infty$.  In this paper, we assume all contact intervals have an absolutely continuous distribution and, for a fixed $i$ or a fixed $j$, the contact intervals $\tau_{ij}$, $i\neq j$, are independent.

\paragraph{Susceptibility and infectiousness processes} Let $S_i(t) = \mathbf{1}_{t\leq t_i}$ and $I_i(t) = \mathbf{1}_{t\in (t_i + \varepsilon_i, t_i + r_i]}$ be the susceptibility and infectiousness processes, respectively, for person $i$, where $\mathbf{1}_X = 1$ if $X$ is true and zero otherwise.  As defined, both processes are left-continuous and infectious contact from $i$ to $j$ is possible at time $t$ only if $C_{ij}I_i(t)S_j(t) = 1$.

\paragraph{Complete observed data} Our population has size $n$, and $m$ represents the number of infections we observe.  Observation begins at time $t = 0$ and ends at time $t = T$.  During this period, we observe the times of all $S\rightarrow E$ (infection), $E\rightarrow I$ (onset of infectiousness), and $I\rightarrow R$ (recovery) transitions that occur in the population.  For all ordered pairs $ij$, we observe $C_{ij}$ and any covariates $X_{ij}$ needed to specify $\lambda_{ij}(\tau)$ up to an unknown parameter vector $\theta$ with true value $\theta_0$.

\subsection{Score processes when who-infects-whom is observed}
Choose an ordered pair $ij$ and let $N_{ij}(t) = \mathbf{1}_{t\geq t_i + \varepsilon_i + \tau_{ij}}$ count the number of infectious contacts from $i$ to $j$ on or before time $t$.  We count only the first infectious contact because $j$ is infected on or before that time.  Consider the filtration
\begin{displaymath}
    \mathcal{H}^{ij}_t = \sigma\big(N_{ij}(u), S_i(u), I_j(u): 0\leq u\leq t\big).
\end{displaymath}
We assume that $N_{ij}(0) = 0$ and $\lambda_{ij}(\tau)$ is predictable with respect to $\mathcal{H}^{ij}_t$, so 
\begin{equation}
    M_{ij}(t) = N_{ij}(t) - \int_0^t \lambda_{ij}(u - t_i - \varepsilon_i)C_{ij}I_i(u)S_j(u)\,du
    \label{Mij}
\end{equation}
is a zero-mean martingale with respect $\mathcal{H}^{ij}_t$.  Now suppose $\lambda_{ij}(\tau)$ is specified up to a parameter vector $\theta$ with true value $\theta_0$, so $\lambda_{ij}(\tau) = \lambda_{ij}(\tau;\theta_0)$.  If the pair $ij$ is observed from time $0$ until time $T$, the corresponding log likelihood is
\begin{displaymath}
    \ell_{ij}(\theta) = \int_0^T \ln\lambda_{ij}(u - t_i - \varepsilon_i;\theta)\,dN_{ij}(u) - \int_0^T \lambda_{ij}(u - t_i - \varepsilon_i;\theta)C_{ij}I_i(t)S_j(u)\,du.
\end{displaymath}
If $\ln\lambda_{ij}(\tau;\theta)$ is differentiable with respect to $\theta$ and we can interchange the order of differentiation and integration, the score process for data in the time interval $[0, t]$ is
\begin{equation}
    U_{ij}(\theta, t) = \int_0^t \frac{\partial}{\partial\theta}\ln\lambda_{ij}(u - t_i - \varepsilon_i;\theta)\,dM_{ij}(\theta, u),
    \label{Uij}
\end{equation}
where 
\begin{displaymath}
    M_{ij}(\theta, u) = N_{ij}(t) - \int_0^t\lambda_{ij}(u - t_i - \varepsilon_i;\theta)C_{ij}I_i(u)S_j(u)\,du.
\end{displaymath}
Therefore, $U_{ij}(\theta_0, t)$ is a zero-mean martingale because it is the integral of a predictable process with respect to $M_{ij}(\theta_0, t)$.  When $C_{ij} = 0$, we have $M_{ij}(\theta, t) = U_{ij}(\theta, t) = 0$ for all $\theta$ and $t$.

Now fix $j$ and assume there exist covariates $X_{ij}$ such that $\lambda_{ij}(\tau;\theta) = \lambda(\tau;\theta,X_{ij})$ for all $i\neq j$.  For each $i\neq j$, assume $N_{ij}(0) = 0$ and $\lambda_{ij}(\tau)$ is predictable with respect to $\mathcal{H}^{ij}_t$.  Since the contact intervals $\tau_{ij}$ are independent for a fixed $j$ and absolutely continuous, the $M_{ij}(\theta_0, \tau)$ from equation \eqref{Mij} are orthogonal zero-mean martingales with respect to the filtration
\begin{displaymath}
    \mathcal{H}^{\cdot j}_t = \sigma\big(N_{ij}(u), I_i(u), S_j(u): 0\leq u\leq t, i\neq j\big).
\end{displaymath}
The total score process for $j$ is
\begin{equation}
    U_{\cdot j}(\theta, t) = \sum_{i\neq j} U_{ij}(\theta, t),
    \label{Uj}
\end{equation}
and $U_{\cdot j}(\theta_0, t)$ is a zero-mean martingale with respect to $\mathcal{H}^{\cdot j}_t$ because it is a sum of zero-mean martingales.  The score process in equation \eqref{Uj} is that of a survival likelihood where the $t_{ij}$ are failure times and $C_{ij}I_i(t)S_j(t) = 1$ indicates risk of infectious contact in the ordered pair $ij$.  At the earliest infectious contact, the contact intervals in all remaining pairs at risk are right-censored, which is a type II independent censoring mechanism \citep{KalbfleischPrentice}.

\subsection{Score processes when who-infects-whom is not observed}
In the previous section, $U_{\cdot j}(\theta, t)$ is adapted only if we observe which of the $N_{ij}(t)$ jumps first, which is equivalent to observing the infector of person $j$.  Now suppose that we observe the infection time of $j$ but not which person $i$ was the infector.  This is equivalent to observing $N_{\cdot j}(t) = \sum_{i\neq j} \int_0^t S_j(u)\,dN_{ij}(u)$, which counts the first infectious contact received by $j$.  The corresponding filtration is 
\begin{displaymath}
    \widetilde{\mathcal{H}}^{\cdot j}_t = \sigma\big(N_{\cdot j}(u), I_i(u), S_j(u): 0\leq u\leq t, i\neq j\big),
\end{displaymath}
and the corresponding zero-mean counting process martingale is $M_{\cdot j}(\theta_0, t)$, where
\begin{equation}
    M_{\cdot j}(\theta, t) = N_{\cdot j}(t) - \int_0^t \lambda_{\cdot j}(u;\theta)S_j(u)\,du
    \label{Mj}
\end{equation}
and $\lambda_{\cdot j}(t;\theta) = \sum_{i\neq j} \lambda(t - t_i - \varepsilon_i;\theta, X_{ij})C_{ij}I_i(t)$.  We can no longer calculate $U_{\cdot j}(\theta, t)$ as defined in equation \eqref{Uj}, but we can calculate its conditional expectation given $\widetilde{\mathcal{H}}^{\cdot j}_t$.  For each $ij$, define the expected score process
\begin{equation}
    \widetilde{U}_{ij}(\theta, t) = \int_0^t \frac{\partial}{\partial\theta}\ln\lambda(u - t_i - \varepsilon_i;\theta, X_{ij}) \,E[dM_{ij}(\theta, u)|\widetilde{\mathcal{H}}^{\cdot j}_t].
    \label{tildeUij}
\end{equation}
Given that $N_{\cdot j}$ jumps at time $t$, the probability that the jump occurred in $N_{ij}$ is
\begin{displaymath}
    \Pr(dN_{ij}(t) = 1|dN_{\cdot j}(t) = 1, \theta, \widetilde{\mathcal{H}}^{\cdot j}_t) = \frac{\lambda(t - t_i - \varepsilon_i;\theta, X_{ij})C_{ij}I_i(t)}{\lambda_{\cdot j}(t;\theta)}.
\end{displaymath}
Thus,
\begin{displaymath}
    E[dM_{ij}(\theta, u)|\widetilde{\mathcal{H}}^{\cdot j}_t] = \frac{\lambda(u - t_i - \varepsilon_i; \theta, X_{ij})C_{ij}I_i(u)}{\lambda_{\cdot j}(u;\theta)}dN_{\cdot j}(u) - \lambda(u - t_i - \varepsilon_i;\theta, X_{ij})C_{ij}I_i(u)S_j(u)\,du,
\end{displaymath}
and equation \eqref{tildeUij} can be rewritten
\begin{displaymath}
    \widetilde{U}_{ij}(\theta, t) = \int_0^t \frac{\frac{\partial}{\partial\theta}\lambda(u - t_i - \varepsilon_i;\theta, X_{ij})C_{ij}I_i(u)}{\lambda_{\cdot j}(u;\theta)}\,dN_{\cdot j}(u) - \int_0^t \frac{\partial}{\partial\theta}\lambda(u - t_i - \varepsilon_i; \theta, X_{ij})C_{ij}I_i(u)S_j(u)\,du.
\end{displaymath}
Therefore, the expected score process for person $j$ is	
\begin{equation}
    \widetilde{U}_{\cdot j}(\theta, t) = \sum_{i\neq j} \widetilde{U}_{ij}(t) = \int_0^t \frac{\partial}{\partial\theta}\ln\lambda_{\cdot j}(u;\theta)\,dM_{\cdot j}(\theta, u),
    \label{tildeUj}
\end{equation}
which is the score process of the of the log likelihood
\begin{equation}
    \widetilde{\ell}_{\cdot j}(\theta) = \int_0^T \ln\lambda_{\cdot j}(u;\theta)\,dN_{\cdot j}(u) - \int_0^T \lambda_{\cdot j}(u;\theta)S_j(u)\,du.
\end{equation}
$\widetilde{U}_{\cdot j}(\theta_0, t)$ is a zero-mean martingale with respect to $\widetilde{\mathcal{H}}^{\cdot j}_t$ because it is the integral of a predictable process with respect to $M_{\cdot j}(\theta_0, t)$.  For an imported infection $j$, $\widetilde{U}_{\cdot j}(\theta, t) = 0$ for all $\theta$ and all $t\in [0, T]$. 

Finally, consider the filtration
\begin{displaymath}
    \widetilde{\mathcal{H}}_t = \sigma\big(N_{\cdot j}(u), I_j(u), S_j(u): 0\leq u\leq t, j = 1,\ldots,n\big)
\end{displaymath}
generated by the complete data described at the end of Section 2.1.  Since we assume that the $\tau_{ij}$, $j\neq i$, are independent for a fixed $i$ and absolutely continuous, the $M_{\cdot j}(\theta_0, t)$ from equation \eqref{Mj} are orthogonal zero-mean martingales with respect to $\widetilde{\mathcal{H}}_t$.  The total expected score process is
\begin{equation}
    \widetilde{U}(\theta, t) = \sum_{j=1}^n \widetilde{U}_{\cdot j}(\theta, t),
    \label{U}
\end{equation}
which is the score process for the log likelihood
\begin{equation}
    \widetilde{\ell}(\theta) = \sum_{j=1}^n \widetilde{\ell}_{\cdot j}(\theta).
    \label{l}
\end{equation}
$\widetilde{U}(\theta_0, t)$ is a zero-mean martingale with respect to $\widetilde{\mathcal{H}}_t$ because it is a sum of zero-mean martingales.  The maximum likelihood estimate (MLE) for $\theta$ is the solution to the equation $\widetilde{U}(\hat{\theta}, T) = 0$.

\subsection{Asymptotic distribution of $\hat{\theta}$}
In this section, we show that the variance of $\widetilde{U}(\theta_0, t)$ can be estimated using its predictable and optional variation processes, which are unbiased estimators of the Fisher information from the survival likelihood.  We then use the Lindeberg-Feller Central Limit Theorem to give a heuristic justification for standard maximum likelihood estimation with epidemic data.  Throughout this section, we assume that $\lambda(\tau;\theta, X)$ has a bounded second derivative in $\theta$ and that integration and differentiation can be interchanged.

Taking the derivative of $U_{\cdot j}(\theta, t)$ with respect to $\theta$ in equation \eqref{tildeUj} leads to 
\begin{displaymath}
    -\frac{\partial}{\partial\theta} \widetilde{U}_{\cdot j}(\theta, t) = \int_0^t \frac{\partial^2}{\partial\theta^2}\ln\lambda_{\cdot j}(u;\theta)\,dM_{\cdot j}(\theta, u) - \int_0^t [\frac{\partial}{\partial\theta}\ln\lambda_{\cdot j}(u;\theta)][\frac{\partial}{\partial\theta}\ln\lambda_{\cdot j}(u;\theta)]^T\lambda_{\cdot j}(u;\theta)S_j(u)\,du.
\end{displaymath}
Setting $\theta = \theta_0$ makes the first term the integral of a predictable process with respect to a zero-mean martingale.  Therefore, 
\begin{equation}
    E\left[-\frac{\partial}{\partial\theta}\widetilde{U}_{\cdot j}(\theta_0, t)\right] = E\left[\int_0^t [\frac{\partial}{\partial\theta}\ln\lambda_{\cdot j}(u;\theta_0)][\frac{\partial}{\partial\theta}\ln\lambda_{\cdot j}(u;\theta_0)]^T\lambda_{\cdot j}(u;\theta_0)S_j(u)\,du\right],
    \label{PredVar} 
\end{equation}
so the predictable variation process $\langle\widetilde{U}_{\cdot j}(\theta_0)\rangle(t)$ is an unbiased estimator of $\text{Var}[\widetilde{U}_{\cdot j}(\theta_0, t)]$.  By equation \eqref{U} and orthogonality of the $\widetilde{U}_{\cdot j}(\theta_0, t)$, the total predictable variation process $\langle\widetilde{U}(\theta_0)\rangle(t) = \sum_j \langle\widetilde{U}_{\cdot j}(\theta_0)\rangle(t)$ is an unbiased estimator of $\text{Var}[\widetilde{U}(\theta_0, t)]$. 

To show that the same result holds for the optional variation process, rearrange equation \eqref{tildeUj} to get
\begin{displaymath}
    \widetilde{U}_{\cdot j}(\theta, t) = \int_0^t \frac{\partial}{\partial\theta}\ln\lambda_{\cdot j}(u;\theta)\,dN_{\cdot j} - \int_0^t \frac{\partial}{\partial\theta}\lambda_{\cdot j}(u;\theta)S_j(u)\,du.
\end{displaymath}
Taking the derivative with respect to $\theta$ yields
\begin{displaymath}
    -\frac{\partial}{\partial\theta} \widetilde{U}_{\cdot j}(\theta, t) = \int_0^t [\frac{\partial}{\partial\theta}\ln\lambda_{\cdot j}(u;\theta)][\frac{\partial}{\partial\theta}\ln\lambda_{\cdot j}(u;\theta)]^T\,dN_{\cdot j}(u) - \int_0^t \frac{\frac{\partial^2}{\partial\theta^2}\lambda_{\cdot j}(u;\theta)}{\lambda_{\cdot j}(u;\theta)}\,dM_{\cdot j}(\theta, u).
\end{displaymath}
Setting $\theta = \theta_0$ makes the second term the integral of a predictable process with respect to a zero-mean martingale.  Therefore,
\begin{equation}
    E\left[-\frac{\partial}{\partial\theta}\widetilde{U}_{\cdot j}(\theta_0, t)\right] = E\left[\int_0^t [\frac{\partial}{\partial\theta}\ln\lambda_{\cdot j}(u;\theta_0)][\frac{\partial}{\partial\theta}\ln\lambda_{\cdot j}(u;\theta_0)]^T\,dN_{\cdot j}(u)\right]
    \label{OptVar}
\end{equation}
so the optional variation process $[\widetilde{U}_{\cdot j}(\theta_0)](t)$ is an unbiased estimator of $\text{Var}[\widetilde{U}_{\cdot j}(\theta_0, t)]$ and the total optional variation process $[\widetilde{U}(\theta_0)](t) = \sum_j [\widetilde{U}_{\cdot j}(\theta_0)](t)$ is an unbiased estimator of $\text{Var}[\widetilde{U}(\theta_0, t)]$.

Imagine a series of epidemics in larger and larger populations, and assume that the final sizes of the epidemics become infinite as the population size $n\rightarrow\infty$.  For any fixed $T$, the number of infections will not become infinite as $n\rightarrow\infty$, which makes it difficult to apply the Martingale Central Limit Theorem to $\widetilde{U}(\theta_0, T)$.  Instead, imagine that we observe $m_n$ infections in a population of size $n$ between time $0$ and time $T_n$, with $m_n\rightarrow\infty$ as $n\rightarrow\infty$.  Let $\widetilde{U}_{n}(\theta, T_{n})$ be the corresponding total expected score process, and let $\hat{\theta}_{n}$ be the corresponding MLE.  If the Lindeberg condition holds for the triangular array $\widetilde{U}_{\cdot 1}(\theta_0, T_n),\ldots,\widetilde{U}_{\cdot n}(\theta_0, T_n)$, then
\begin{displaymath}
    \frac{\widetilde{U}_{n}(\theta_0, T_n)}{\text{Var}[\widetilde{U}_{n}(\theta_0, T_n)]^\frac{1}{2}} \;\longrightarrow\; N(0, 1)
\end{displaymath} 
in distribution as $n\rightarrow\infty$ by the Lindeberg-Feller Central Limit Theorem \citep{Serfling}.  Heuristically, this justifies the use of maximum likelihood methods such as Wald, score, and likelihood ratio tests.

\section{Estimation of $R_0$}
The contact interval distribution can be used to estimate $R_0$ in both network-based and mass-action models.  For simplicity, we assume that the hazard of infectious contact does not depend on covariates.  Thus, $\lambda_{ij}(\tau;\theta) = \lambda(\tau;\theta)$ for all $ij$ and the results in this section apply to homogeneous populations.  For mass-action models, we describe an asymptotic likelihood that is valid for the initial spread of disease.

\subsection{Network-based models}
In a network-based model, transmission takes place across the edges of a \textit{contact network}, so we have $C_{ij} = 1$ if and only if there is an edge leading from $i$ to $j$ in the contact network.  Here, we will assume that contact networks are undirected, so $C_{ij} = C_{ji}$ for all $i$ and $j$.  In a network-based model, $R_0$ depends on the structure of the contact network.  The most tractable models are those on \textit{configuration-model} networks, which are maximally random except for their degree distribution \citep{MolloyReed1, MolloyReed2, NSW}.  More formally, let $D$ be a nonnegative discrete random variable with finite mean and variance.  To construct a configuration-model network with $n$ nodes, assign each node $i = 1,\ldots, n$ a degree $d_i$ randomly sampled from the distribution of $D$.  Then connect the stubs at random, erasing one stub if necessary so the sum of the degrees is even.  As $n\rightarrow\infty$, the probability of multiple edges between two nodes or loop from a node to itself goes to zero.  

In these networks, there is a straighforward definition of $R_0$ \citep{Andersson, Newman1, Kenah1}.  In the early stages of transmission, an infected node of degree $d$ has $d-1$ edges across which infection can be transmitted.  The probability of transmitting infection across each of these edges is $\exp(-\Lambda(\iota;\theta_0))$, where $\iota$ is the infectious period and $\Lambda(t;\theta) = \int_0^t \lambda(u;\theta)\,du$.  Since the probability of reaching a node by following edges is proportional to the degree of the node, the mean number of secondary infections generated by a typical infected node in the early stages of an epidemic is
\begin{equation}
    R_0 = E[e^{-\Lambda(\iota;\theta_0)}]\Big(\frac{E[D^2]}{E[D]} - 1\Big),
    \label{netR0}
\end{equation}
where the first expectation is taken over the distribution of the infectious period $\iota$.  

\paragraph{Network-based likelihood} In a network-based model, the likelihood $\widetilde{\ell}(\theta)$ in equation \eqref{l} depends only on data about individuals who are either infected before time $T$ or connected to an infected person in the contact network.  In principle, these people could be identified through surveillance and contact tracing.  For all other individuals $j$, $\widetilde{U}_{\cdot j}(\theta, t) = 0$ for all $t\in [0, T]$ because $C_{ij}I_i(t) = 0$ for all $i$.  Since $\frac{E[D^2]}{E[D]}$ is the expected degree of persons who are infected by transmission within the population, it can be estimated by calculating the mean degree of persons who are infected.

\subsection{Mass-action models}
In a mass-action model, individuals form no stable social bonds and interact like gas molecules.  Thus, $C_{ij} = 1$ for all $ij$ but the hazard of infectious contact is inversely proportional to the population size.  If $\lambda_{n}(\tau;\theta)$ is the hazard function for the contact interval distribution in a population of size $n$,
\begin{displaymath}
    \lambda_{n}(\tau;\theta) = \frac{\lambda_0(\tau;\theta)}{n-1}
\end{displaymath}
for a \textit{baseline hazard function} $\lambda_0(\tau;\theta)$ with corresponding cumulative hazard function $\Lambda_0(\tau;\theta)$.  As before, these functions are specified up to an unknown parameter vector $\theta$ with true value $\theta_0$.

The baseline hazard and cumulative hazard functions of a mass-action model have useful interpretations in terms of $R_0$ and the time course of infectiousness in the limit as $n\rightarrow\infty$.  Given an infectious period $\iota$, the expected number of infectious contacts made is
\begin{equation}
    R_0 = (n-1)\left(1 - e^{-\frac{1}{n-1}\Lambda_0(\iota;\theta_0)}\right) \;\longrightarrow\; \Lambda_0(\iota;\theta_0).
    \label{maR0}
\end{equation}
Given that $i$ makes infectious contact with $j$ and has infectious period $\iota$, the probability density function of the infectious contact interval from $i$ to $j$ is
\begin{equation}
    \frac{\frac{1}{n-1}\lambda_0(\tau;\theta)e^{-\frac{1}{n-1}\Lambda_0(\tau;\theta_0)}}{1 - e^{-\frac{1}{n-1}\Lambda_0(\iota;\theta_0)}} \;\longrightarrow\; \frac{\lambda_0(\tau;\theta_0)}{R_0}.
\end{equation}

\paragraph{Mass-action likelihood} Let $m$ be the total number of infections observed before time $T$.  If $m\ll n$, an approximate likelihood that depends only on information about infected presons can be written in terms of $\lambda_0(\tau;\theta)$.  Expanding equation \eqref{l} in terms of $\lambda_0(\tau;\theta)$, we get
\begin{align}
    \widetilde{\ell}(\theta) &= \sum_{j=1}^n \int_0^T \ln\Big(\sum_{i\neq j} \lambda_0(u - t_i - \varepsilon_i;\theta)I_i(u)\Big)\,dN_{\cdot j}(u) - \sum_{j=1}^n\int_0^T \ln(n-1)\,dN_{\cdot j}(u) \nonumber \\
    &\qquad - \frac{1}{n-1}\sum_{j=1}^n \int_0^T (\sum_{i\neq j}\lambda_0(u - t_i - \varepsilon_i;\theta)I_i(u))S_j(u)\,du.
\end{align}
All summands in the first term are zero except for those $j$ with $t_j\leq T$.  The second term is not a function of $\theta$ and can be ignored.  The third term can be split into terms from $j$ who get infected on or before time $T$ and from those who remain uninfected at time $T$:
\begin{displaymath}
    \frac{1}{n-1}\sum_{j:t_j\leq T} \bigg(\sum_{i:t_i<t_j} \Lambda_0\big((t_j - t_i - \varepsilon_i)\wedge\iota_i;\theta\big)\bigg) + \frac{n-m}{n-1}\sum_{i:t_i\leq T} \Lambda_0\big((T - t_i - \varepsilon_i)\wedge\iota_i;\theta\big),
\end{displaymath}
where $x\wedge y = \min(x, y)$.  Since the first term is less than or equal to 
\begin{displaymath}
    \frac{m}{n-1}\sum_{i:t_i\leq T}\Lambda_0\big((T - t_i - \varepsilon_i)\wedge\iota_i;\theta\big),
\end{displaymath}
we have
\begin{equation}
    \tilde{\ell}(\theta) \;\longrightarrow\; \sum_{j:t_j\leq T} \bigg(\ln\Big(\sum_{i\neq j} \lambda_0(t_j - t_i - \varepsilon_i;\theta)I_i(t_j)\Big) - \Lambda_0\big((T - t_j - \varepsilon_j)\wedge \iota_j;\theta\big)\bigg)
    \label{asympl}
\end{equation}
for a fixed $m$ as $n\rightarrow\infty$.  This asymptotic likelihood depends only on information about infected people.  In principle, these people could be identified through surveillance.

\section{Simulations and illustration}
In this section, we first look at the performance of the methods from Sections 2 and 3 in simulated epidemic data sets from mass-action and network-based models.  We then illustrate the use of our methods with an analysis of two epidemic curves from the early spread of influenza A(H1N1) in Mexico.

\subsection{Simulations}
In this section, we look at the performance of the methods from Sections 2 and 3 in simulated epidemic data sets.  In all models, we used data from the first $m = 1,000$ infections in a population of size $n=100,000$.  For each infected person $i$, we recorded the infection time $t_i$, the onset of infectiousness $t_i+\epsilon_i$, and the recovery time $t_i + r_i$.  In network-based models, the degree $d_i$  and the indices of all neighbors of $i$ were also recorded.  All outbreaks started with a single imported infection at time $0$.  Since we are interested primarily in the analysis of emerging epidemics, outbreaks that terminated with a final size less than 1,000 were discarded.  If an epidemic model was run 100 times without producing an epidemic final size of at least 1,000, it was discarded and another model was generated.  For all simulations, $R_0$ was constrained to be between 1.01 and 16, a range that covers almost all known epidemic diseases.

In network-based models, the contact networks were undirected Erd\H{o}s-R\'{e}nyi random graphs \citep{NetworksBook} with an expected degree chosen from the discrete uniform distribution on $\{2,\ldots, 16\}$.  A new contact network was constructed for each simulation.  

Four scenarios were considered within each class of model: exponential or Weibull (baseline) contact interval distributions with constant or exponentially-distributed infectious periods.  All infectious period distributions had mean one.  The exponential distribution has the hazard function $\lambda(\tau;\beta) = \beta$ for all $\tau > 0$, where $\beta > 0$ is the rate parameter.  The Weibull distribution has the hazard function $\lambda(\tau;\alpha, \beta) = \alpha\beta(\beta\tau)^{\alpha-1}$ for all $\tau > 0$, where $\alpha > 0$ is the shape parameter and $\beta > 0$ is the rate parameter.  Note that the exponential distribution is a Weibull distribution with $\alpha=1$.

\paragraph{Parameter estimates} For network-based models, we used the likelihood in equation \eqref{l} to estimate the parameters of the contact interval distribution.  For mass-action models, we used the asymptotic likelihood in equation \eqref{asympl} to estimate the parameters of the baseline contact interval distribution.  Maximum likelihood estimates were obtained using the \texttt{mle} function in the R library \texttt{stats4}.  Confidence intervals for each parameter were calculated using the \texttt{confint} function, which inverts the one-parameter likelihood ratio chi-squared test using a profile likelihood.

\paragraph{$R_0$ estimates} For network-based models, $R_0$ was estimated using equation \eqref{netR0}.  For mass-action models, $R_0$ was estimated using equation \eqref{maR0}.  We calculated bootstrap percentile confidence intervals by sampling contact interval distribution parameters from their approximate joint normal distribution and combining each sample with a bootstrap sample of the observed infectious periods (and, for network-based models, observed degrees in the contact network).  The 95\% confidence interval was defined by the 2.5\% and 97.5\% quantiles of the point estimates from 10,000 samples.

\paragraph{Implementation} Simulations were implemented in Python 2.6 (www.python.org) using the SciPy 0.7 package \citep{SciPy}.  Analyses were performed in R 2.10 \citep{R} via the RPy 2.0 package \citep{RPy}.  Contact networks were generated using the NetworkX 0.99 package \citep{NetworkX}.  Sampling from multivariate normal distributions was done using the Cholesky distribution of the covariance matrix \citep{Rizzo}.  The simulation code is included as supplementary material (http://www.biostatistics.oxfordjournals.org).

\subsubsection{Mass-action models}
For mass-action models with exponential contact intervals, $R_0 = \beta$ for both fixed and exponentially-distributed infectious periods.  Let $\hat{\beta}$ denote the MLE of the rate parameter $\beta$, and let $\iota_k$ denote the infectious period of the $k^\text{th}$ infection observed.  Our point estimate of $R_0$ is
\begin{equation}
    \hat{R}_0 = \frac{1}{m}\sum_{k=1}^m\hat{\beta}\iota_i.
    \label{mA_expR0}
\end{equation}
A bootstrap sample of $R_0$ is 
\begin{equation}
    R^*_0 = \frac{1}{m}\sum_{k=1}^m\beta^*\iota^*_k,
\end{equation}
where $\beta^*$ is a parametric bootstrap sample from the approximate normal distribution of $\hat{\beta}$ and $\iota^*_1,\ldots, \iota^*_m$ is a bootstrap sample from the observed $\iota_1,\ldots, \iota_m$.

For mass-action models with Weibull contact intervals $R_0 = \beta^\alpha$ for a fixed infectious period and $R_0 = \beta^\alpha\Gamma(\alpha+1)$ for exponentially-distributed infectious periods.  In both cases, 
\begin{equation}
    \hat{R}_0 = \frac{1}{m}\sum_{k=1}^m(\hat{\beta}\iota_k)^{\hat{\alpha}},
    \label{mA_WeibR0}
\end{equation}
where $\hat{\alpha}$ is the shape parameter MLE and $\hat{\beta}$ is the rate parameter MLE.  A bootstrap sample of $R_0$ is
\begin{equation}
    R^*_0 = \frac{1}{m}\sum_{k=1}^m(\beta^*\iota^*_k)^{\alpha^*},
\end{equation}
where $(\alpha^*, \beta^*)$ is a sample from the approximate joint normal distribution of $(\hat{\alpha}, \hat{\beta})$.

\begin{figure}
    \includegraphics[width=\textwidth]{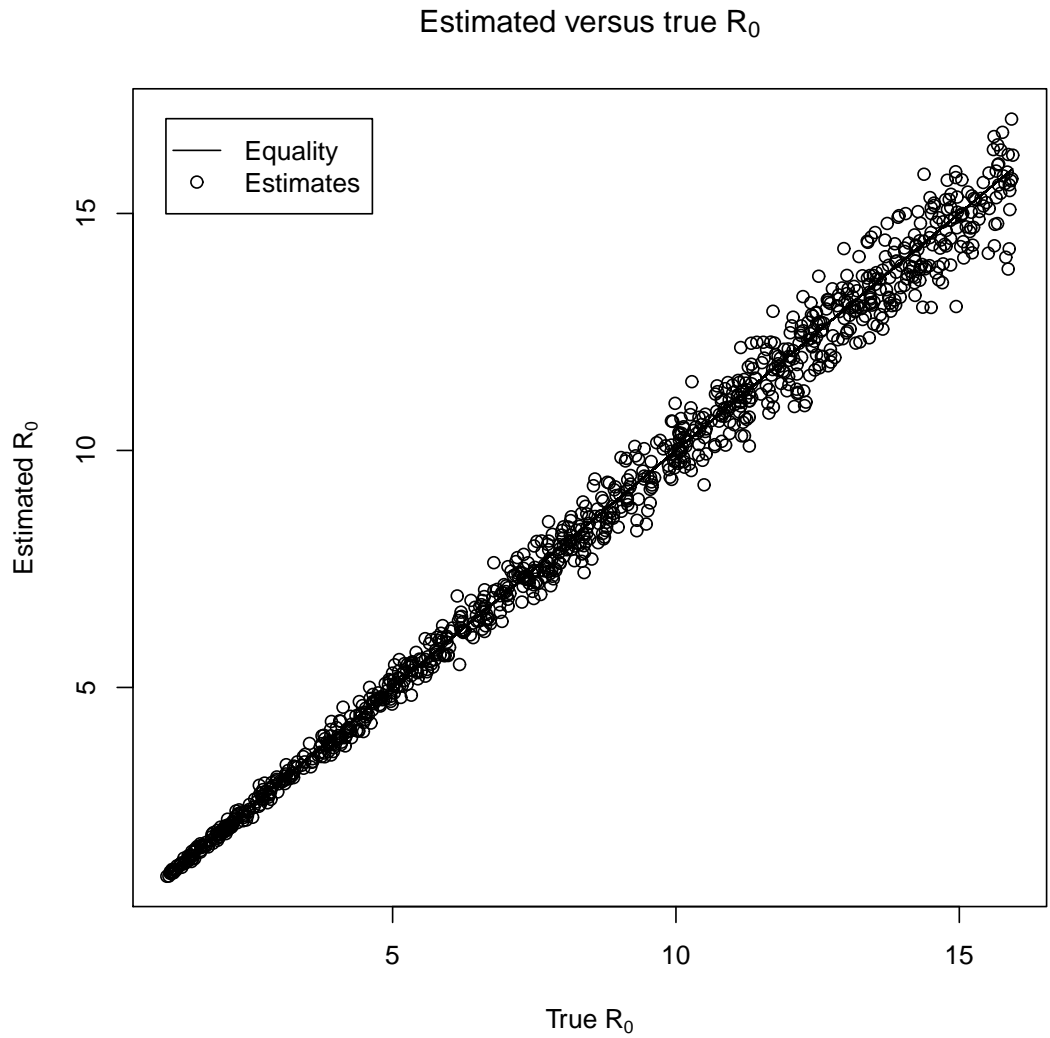}
    \caption{\label{expCIexpIPr0_mA} Scatterplot of estimated versus true $R_0$ for mass-action models with exponential contact interval and infectious period distributions, showing excellent agreement.  Similar results were obtained in models with a fixed infectious period (not shown).}
\end{figure}

\begin{figure}
    \includegraphics[width=\textwidth]{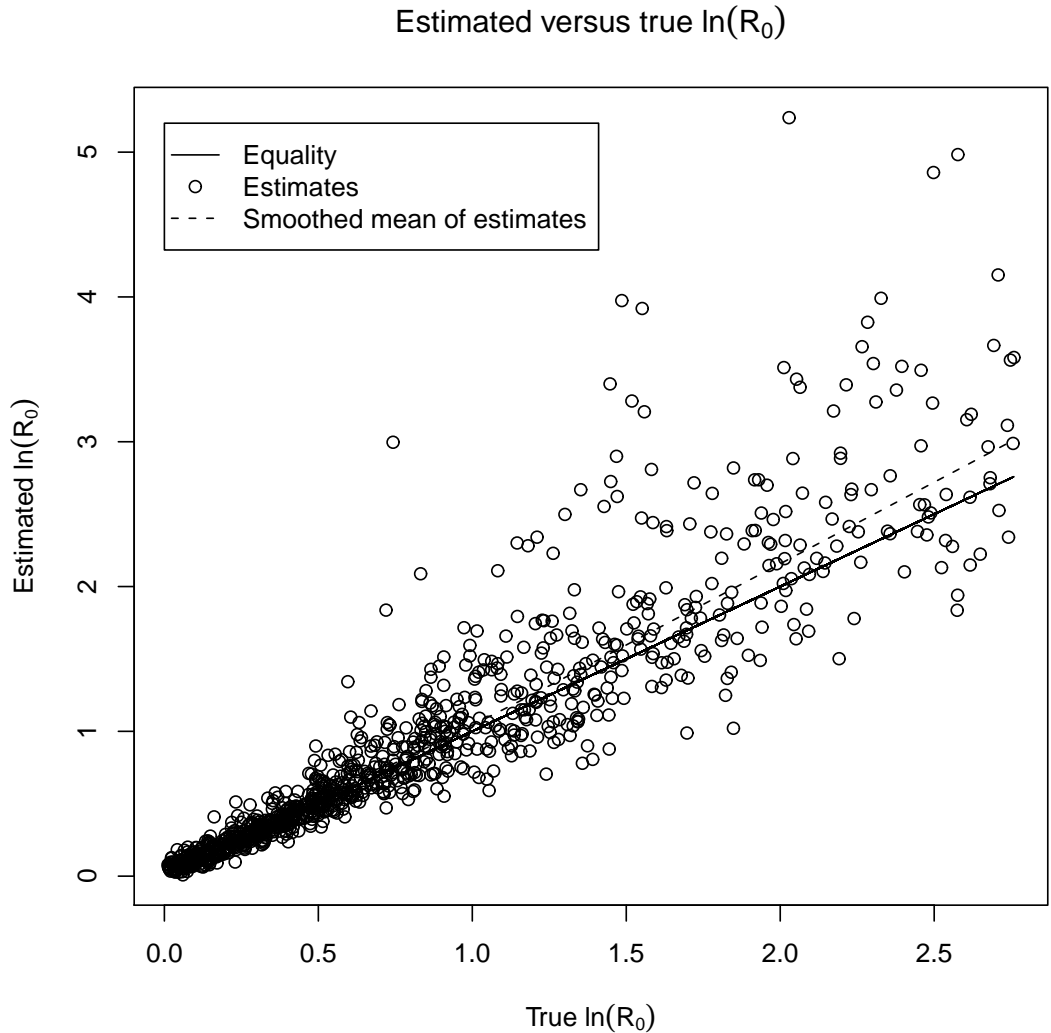}
    \caption{\label{WeibCIexpIPlnR0_mA} Scatterplot of estimated versus true $\ln(R_0)$ for mass-action models with Weibull contact interval distributions and exponential infectious period distributions.  Estimates are nearly unbiased at low $R_0$, but biased upward at high $R_0$.  Similar results were obtained in models with a fixed infectious period (not shown).  The smoothed mean was produced with the R command \texttt{lowess}.  One simulation that produced $\hat{R}_0=3730.8$ (1089.9, 12,245.8) was excluded from the graph; it had a true $R_0 = 10.5$.}
\end{figure}

\paragraph{Results} Table \ref{massActionTable} shows the coverage probabilities achieved in 1,000 simulations and exact binomial 95\% confidence intervals for the true coverage probabilities in each of the four types of mass-action model.  Figure \ref{expCIexpIPr0_mA} shows a scatterplot of $\hat{R}_0$ versus $R_0$ for models with exponential contact interval and infectious period distributions.  Figure \ref{WeibCIexpIPlnR0_mA} shows a scatterplot of estimated versus true $\ln(R_0)$ for models with Weibull contact interval distributions and exponential infectious period distributions.  For these models, estimates of $R_0$ are right-skewed because of exponent $\hat{\alpha}$ in equation \eqref{mA_WeibR0}; this is reduced by taking logarithms.  Similar results were obtained in models with a fixed infectious period. 

\subsubsection{Network-based models}
Let $\iota_k$ and $d_k$ denote the infectious period and degree, respectively, of the $k^\text{th}$ infection observed.  In a contact network with $n$ nodes, let $\bar{D}$ be the mean degree and let
\begin{displaymath}
    \widetilde{D} = \bar{D}^{-1}\sum_{i=1}^n d_i(d_i - 1).
\end{displaymath}
For network-based models with exponential contact intervals, $R_0 = (1-\exp(-\beta))\widetilde{D}$ for a fixed infectious period and $R_0 = \frac{\lambda}{\lambda+1}\widetilde{D}$ for exponentially-distributed infectious periods.  In both cases,
\begin{displaymath}
    \hat{R}_0 = \frac{1}{m}\sum_{k=1}^m (1-e^{-\hat{\beta}\iota_k})(d_k - 1).
\end{displaymath}
A bootstrap sample of $R_0$ is
\begin{displaymath}
    R^*_0 = \frac{1}{m}\sum_{k=1}^m (1-e^{-\beta^*\iota_k^*})(d_k^* - 1),
\end{displaymath}
where $\beta^*$ is a sample from the approximate normal distribution of $\hat{\beta}$ and $(\iota^*_1, d^*_1),\ldots, (\iota^*_m, d^*_m)$ is a bootstrap sample from $(\iota_1, d_1),\ldots, (\iota_m, d_m)$.

For network-based models with Weibull contact intervals, $R_0 = (1 - \exp(-\beta^\alpha))\widetilde{D}$ for a fixed infectious period and
\begin{displaymath}
    R_0 = 1 - \int_0^\infty e^{-(\beta x)^\alpha - x}\,dx.
\end{displaymath}
for exponentially-distributed infectious periods.  In both cases,
\begin{displaymath}
    \hat{R}_0 = \frac{1}{m}\sum_{k=1}^m (1-e^{-(\hat{\beta}\iota_k)^{\hat{\alpha}}})(d_k - 1).
\end{displaymath}
A bootstrap sample of $R_0$ is
\begin{displaymath}
    R^*_0 = \frac{1}{m}\sum_{k=1}^m (1-e^{-(\beta^*\iota_k^*)^{\alpha^*}})(d_k^* - 1),
\end{displaymath}
where $(\alpha^*, \beta^*)$ is a sample from the approximate joint normal distribution of $(\hat{\alpha}, \hat{\beta})$ and $(\iota^*_1, d^*_1),\ldots, (\iota^*_m, d^*_m)$ is a bootstrap sample from $(\iota_1, d_1),\ldots, (\iota_m, d_m)$.

\begin{figure}
    \includegraphics[width=\textwidth]{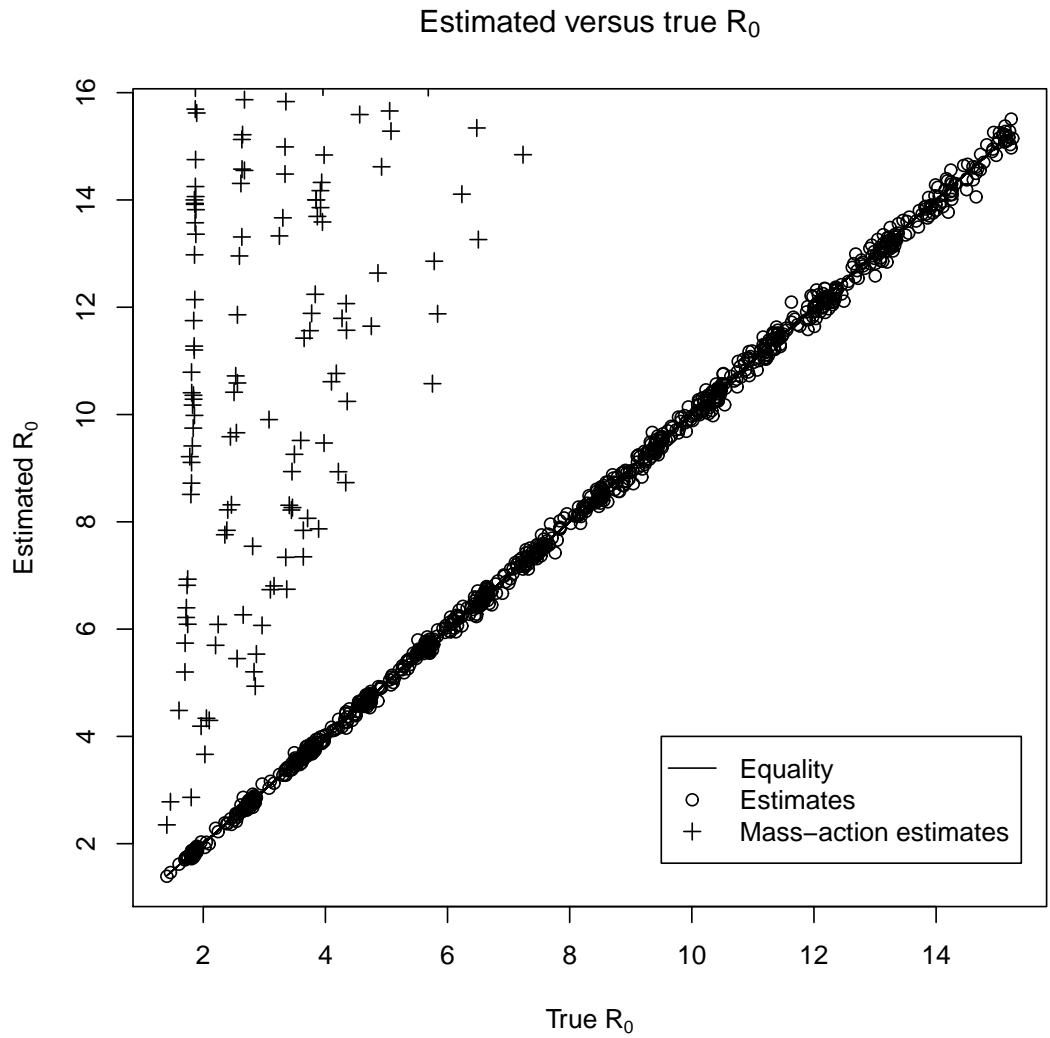}
    \caption{\label{expCIexpIPr0_net} Scatterplot of estimated versus true $R_0$ for network-based models with exponential contact interval and infectious period distributions, showing excellent agreement.  The mass-action estimates are severely biased upward; most are out of range of the plot.  Similar results were obtained in models with a fixed infectious period (not shown).}
\end{figure}

\begin{figure}
    \includegraphics[width=\textwidth]{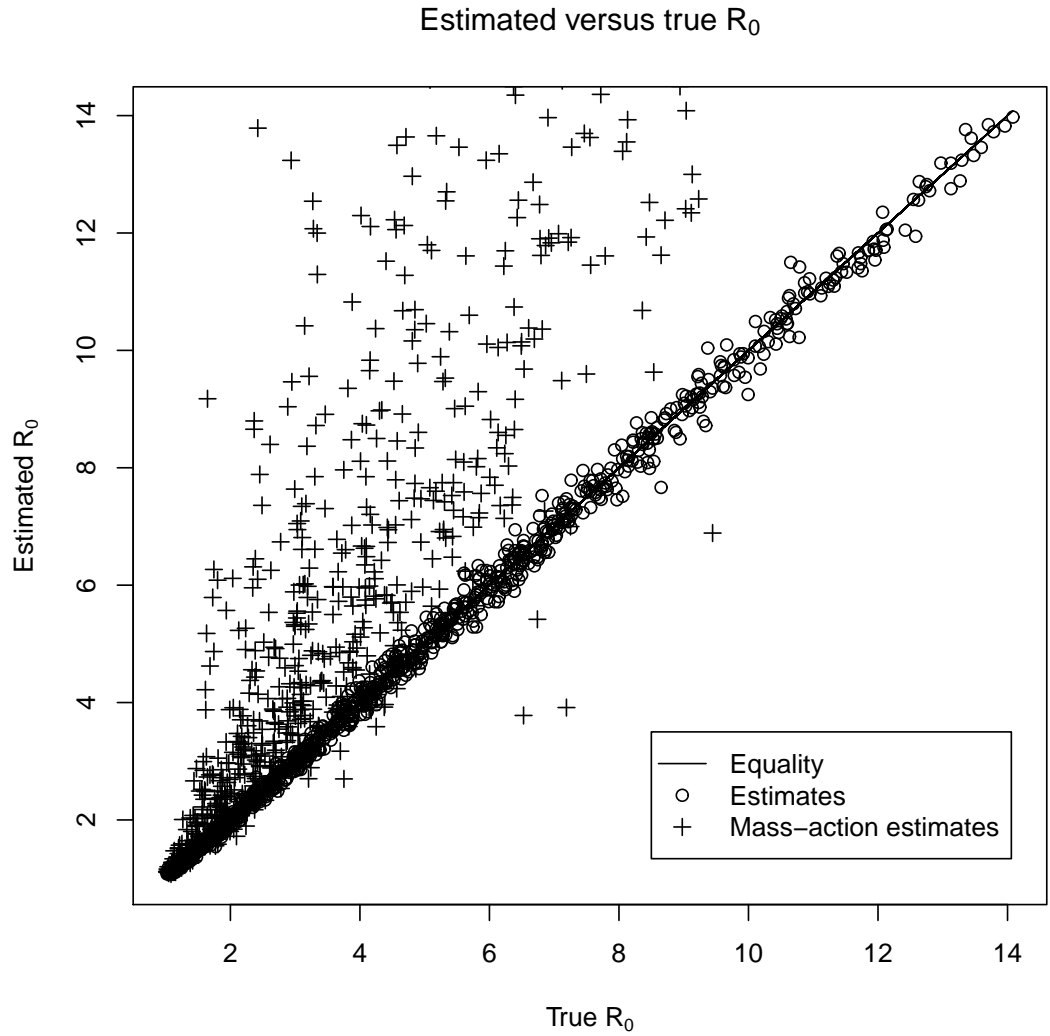}
    \caption{\label{WeibCIexpIPr0_net} Scatterplot of estimated versus true $R_0$ for network-based models with Weibull contact interval distributions and exponential infectious period distributions, showing excellent agreement.  The mass-action estimates are severely biased upward.  Similar results were obtained in models with a fixed infectious period (not shown).}
\end{figure}

\paragraph{Results} Table \ref{networkTable} shows the coverage probabilities achieved in 1,000 simulations and exact binomial 95\% confidence intervals for the true coverage probability in each of the four types of network-based model.  Figure \ref{expCIexpIPr0_net} shows a scatterplot of the estimated versus true $R_0$ for models with exponential contact interval and infectious period distributions.  Figure \ref{WeibCIexpIPr0_net} shows a scatterplot of the estimated versus true $R_0$ for models with Weibull contact interval distributions and exponential infectious period distributions.  Similar results were obtained in models with a fixed infectious period.

\paragraph{Mass-action estimates} To look at the effect of assumptions about the contact process on statistical inference during an epidemic, we applied the mass-action likelihoods to data generated by the network-based models, ignoring all information about the contact network.  Table \ref{networkTable} shows the coverage probabilities achieved in 1,000 simulations and exact binomial 95\% confidence intervals for the true coverage probabilities for mass-action estimates applied to network-based models.  The `+' signs in Figures \ref{expCIexpIPr0_net} and \ref{WeibCIexpIPr0_net} show the mass action estimates of $R_0$ versus the true $R_0$ in network-based models with exponential infectious periods.  Many of these points fall above the top edge of each graph.  Similar results were obtained in models with a fixed infectious period.

\subsection{Illustration: Influenza A(H1N1) in Mexico, 2009}
To show the practicability of methods based on contact intervals as well as the importance of data that is uncollected or unreported in emerging epidemics, we attempted to estimate $R_0$ based on two epidemic curves published at the beginning of the influenza A(H1N1) pandemic in Mexico.  The first epidemic curve contains suspected cases in the village of Vera Cruz between March 9 and March 20 \citep{FraserH1N1}.  The second epidemic curve contains lab-confirmed cases in Mexico City between April 13 and April 24 \citep{MexicoMH}.  In both analyses, we assumed a latent period (between infection and the onset of infectiousness) of one day and an incubation period (between infection and onset of symptoms) of two days.  With no data on links between cases or the duration of illness in each case, we assumed mass-action with a constant infectious period.  Confidence intervals are generated as in the simulations.  

Assuming an exponential contact interval distribution, we get $\hat{R}_0 = 1.95\;(1.63, 2.33)$ for Vera Cruz and $\hat{R}_0 = 2.31\;(2.15, 2.48)$ for Mexico City.  These are high but consistent with some early estimates \citep{FraserH1N1, YangH1N1}.  Assuming a Weibull contact interval distribution, we get $\hat{R}_0 = 3.08\;(2.55, 3.65)$ for Vera Cruz and $\hat{R}_0 = 4.37\;(4.06, 4.70)$ for Mexico City; in both cases, the null hypothesis of an exponential contact interval distribution is strongly rejected (likelihood ratio p-value $<.001$).  The estimates are also sensitive to the assumed infectious period.  Assuming a five-day infectious period and a Weibull contact interval distribution, we get $\hat{R}_0 = 3.53\;(2.79, 4.30)$ for Vera Cruz and $\hat{R}_0 = 7.14\;(6.63, 7.66)$ for Mexico City.  Subsequent experience shows that these $R_0$ estimates are far too high.  This bias is consistent with the results obtained above when applying mass-action estimates to simulated data generated by network-based models.  Since a most influenza transmission takes place in households, workplaces, and schools \citep{YangH1N1} the true underlying transmission model is probably closer to a network-based model than a mass-action model.  Data on the duration of illness and, more importantly, on the social links between cases would allow better point and interval estimates of $R_0$.  The estimates could also be improved with incomplete-data methods that took into account the discreteness of the data and allowed variability in latent, incubation, and infectious periods.

\section{Discussion}
The results of the simulations confirm that standard maximum-likelihood methods can be applied successfully to survival likelihoods written in terms of the contact interval distribution.  In the mass-action models, performance deteriorated noticeably in moving from exponential to Weibull contact interval distributions, possibly because $\widetilde{U}(\theta_0, T)$ was closer to a normal distribution in the simpler models.  No such deterioration was noticeable in the network-based models, possibly due to the addition of contact-tracing information.  Our methods were deliberately simple: all point estimates were plug-in estimators and all confidence intervals were based on normal approximations for the joint distributions of the MLEs.  More sophisticated methods, such as Bayesian methods, might produce point estimates and confidence intervals whose performance is even better.  The methods here would adapt quite well to a Bayesian analysis, and we believe that a Bayesian framework is the most natural setting for the development of methods to analyze partially-observed epidemics.  

Methods based on contact intervals can incorporate a much greater variety of transmission models than methods based on generation or serial intervals, which usually assume mass-action.  The simulation results presented above show that this flexibility is essential for accurate statistical inference during an epidemic.  The mass action estimates failed spectacularly when applied to data generated by network-based models.  The point estimates were severely biased upward, and all 95\% confidence intervals had coverage probabilities below 85\%, with most below 25\%. 

The methods and simulation results in this paper have important implications for data collection during an emerging epidemic.  First, they require information on the onset and duration of infectiousness.  For an acute infectious disease, the onset and duration of illness may provide a useful proxy, especially if there is some knowledge of the incubation period and the pattern of pathogen shedding.  Second, they show the potential value of data about close contacts of cases, whether or not they are infected.  Methods based on generation and serial intervals do not require such data, but this apparent advantage comes at a tremendous cost in terms of the flexibility and validity of the subsequent analysis.  They are essentially missing-data methods with no complete-data counterparts, and they almost certainly understate the true data requirements for accurate estimation of $R_0$.

\paragraph{Limitations} The SEIR framework limits our methods to acute, immunizing diseases that spread person-to-person.  It does not apply to many diseases of public health importance, such as tuberculosis, meningococcal or pneumococcal diseases, foodborne or waterborne diseases, or HIV/AIDS.  Most (though not all) emerging infections fit into the SEIR framework, and almost all methods currently used to analyze data from emerging epidemics make this assumption.  We also assumed that all times of infection, onset of infectiousness, and recovery are observed.  This is clearly unsatisfactory, but the development of incomplete-data methods must be based on complete-data methods.  In Section 2, we assumed that the contact interval $\tau_{ij}$ is independent of the infectious period $\iota_i$ of $i$.  This simplified the likelihoods, but it is probably unrealistic.  This problem could be addressed by including $\iota_i$ as a covariate in $X_{ij}$ or by using multivariate survival methods.  In Section 3, we assumed that the population is homogeneous.  This simplified the estimation of $R_0$, but it is also unrealistic.  In a heterogeneous population, estimates of $R_0$ would have to include the distribution of relevant covariates in the population.

Despite these limitations, methods based on contact intervals and survival analysis have the potential to become important tools in infectious disease epidemiology.  The purpose of this paper was to introduce survival analysis based on contact intervals as a useful complete-data method, and we have done so in the simplest setting possible.  These methods can be seen as descendants of methods based on generation and serial intervals, but they are more flexible and more explicit about assumptions and data requirements.

\vspace*{-6pt}
\section*{Acknowledgments}
I would like to thank M. Elizabeth Halloran for her guidance throughout the preparation this manuscript.  I am also grateful for the comments of Yang Yang, Ira M. Longini, Jr., participants in the workshop ``Design and Analysis of Infectious Disease Studies'' (Mathematisches Forschungsinstitut Oberwolfach, 1-7 November 2009), and the anonymous referees of \textit{Biostatistics}.  This work was supported by National Institute of General Medical Sciences grant F32GM085945, ``Linking transmission models and data analysis in infectious disease epidemiology''. Office space and administrative support were provided by the Fred Hutchinson Cancer Research Center.  
{\textit{Conflict of interest:} None declared.}

\vspace*{-6pt}
\bibliographystyle{chicago}
\bibliography{pCIestimation}

\begin{table}[p]
    \caption{\label{massActionTable} Coverage probabilities for mass-action models.}
    \centering
    \fbox{
    \begin{tabular}{cccc}
        Infectious period & & & \\
        distribution & Parameter & Coverage probability & Exact binomial 95\% CI \\
        \hline
        & \multicolumn{3}{c}{\textbf{Exponential contact interval}} \\
        & $\beta$ & .952 & (.937, .964) \\
        \raisebox{1.5ex}[0pt]{\textbf{Constant}} & $R_0$ & .950 & (.935, .962) \\
        & & & \\
        & $\beta$ & .951 & (.936, .964) \\
        \raisebox{1.5ex}[0pt]{\textbf{Exponential}} & $R_0$ & .963 & (.949, .974) \\
        & & & \\
        & \multicolumn{3}{c}{\textbf{Weibull contact interval}}\\
        & $\alpha$ & .936 & (.919, .950) \\
        \textbf{Constant} & $\beta$ & .907 & (.887, .924) \\
        & $R_0$ & .879 & (.857, .899) \\
        & & & \\
        & $\alpha$ & .927 & (.909, .942) \\
        \textbf{Exponential} & $\beta$ & .912 & (.893, .929) \\
        & $R_0$ & .902 & (.882, .920) \\
    \end{tabular}
    }
\end{table}

\begin{table}[p]
    \caption{\label{networkTable} Coverage probabilities for network-based models.}
    \centering
    \fbox{
    \begin{tabular}{cccccc}
        & & \multicolumn{2}{c}{\underline{Network-based estimates}} & \multicolumn{2}{c}{\underline{Mass-action estimates}} \\
        Infectious period & & Coverage & Exact binomial & Coverage & Exact binomial \\
        distribution & Parameter & probability & 95\% CI & probability & 95\% CI \\
        \hline
        & \multicolumn{5}{c}{\textbf{Exponential contact interval}} \\
        & $\beta$ & .942 & (.926, .956) & .004 & (.001, .010) \\
        \raisebox{1.5ex}[0pt]{\textbf{Constant}} & $R_0$ & .948 & (.932, .961) & .210 & (.185, .237) \\
        & & & & & \\
        & $\beta$ & .943 & (.927, .957) & .035 & (.024, .048) \\
        \raisebox{1.5ex}[0pt]{\textbf{Exponential}} & $R_0$ & .962 & (.948, .973) & .000 & (.000, .004) \\
        & & & & & \\
        & \multicolumn{5}{c}{\textbf{Weibull contact interval}}\\
        & $\alpha$ & .936 & (.919, .950) & .798 & (.772, .822) \\
        \textbf{Constant} & $\beta$ & .946 & (.930, .959) & .025 & (.016, .037) \\
        & $R_0$ & .945 & (.929, .958) & .509 & (.477, .540) \\
        & & & & & \\
        & $\alpha$ & .946 & (.930, .959) & .834 & (.809, .857) \\
        \textbf{Exponential} & $\beta$ & .950 & (.935, .963) & .031 & (.021, .044) \\
        & $R_0$ & .941 & (.925, .955) & .392 & (.362, .423) \\
    \end{tabular}
    }
\end{table}

\end{document}